\documentclass[%
 aip,
 amsmath,amssymb,
 reprint,%
]{revtex4-1}

\usepackage{caption}
\usepackage{subcaption}
\newcommand{\beq}{\begin{equation}}
\newcommand{\eeq}{\end{equation}}
\usepackage{braket}
\usepackage{array}
\usepackage{mathtools, amsmath,amssymb,amsfonts}
\usepackage{graphicx}
\usepackage{dcolumn}
\usepackage{bm}
\usepackage{ragged2e}

\usepackage{xcolor}

\usepackage[utf8]{inputenc}
\usepackage[T1]{fontenc}
\usepackage{mathptmx}
\usepackage{etoolbox}

\makeatletter
\def\@email#1#2{%
 \endgroup
 \patchcmd{\titleblock@produce}
  {\frontmatter@RRAPformat}
  {\frontmatter@RRAPformat{\produce@RRAP{*#1\href{mailto:#2}{#2}}}\frontmatter@RRAPformat}
  {}{}
}%

\makeatother
\begin{document}

\preprint{AIP/123-QED}

\title{Nanoscale single-electron box with a floating lead for quantum sensing: modelling and device characterization}
\author{N. Petropoulos}
\altaffiliation{Authors to whom correspondence should be addressed: nikolaos.petropoulos@ucdconnect.ie and elena.blokhina@ucd.ie}

\affiliation{Centre for Quantum Engineering, Science, and Technology (C-QuEST), Belfield, Ireland}

\affiliation{School of Electrical and Electronic Engineering, University College Dublin, Belfield, Ireland}

\affiliation{Equal1 Labs, Nexus UCD, Belfield, Ireland}

\author{X. Wu}

\affiliation{Centre for Quantum Engineering, Science, and Technology (C-QuEST), Belfield, Ireland}

\affiliation{School of Electrical and Electronic Engineering, University College Dublin, Belfield, Ireland}

\affiliation{Equal1 Labs, Nexus UCD, Belfield, Ireland}

\author{A. Sokolov}

\affiliation{Equal1 Labs, Nexus UCD, Belfield, Ireland}

\author{P. Giounanlis}

\affiliation{Equal1 Labs, Nexus UCD, Belfield, Ireland}
\author{I. Bashir}

\affiliation{Equal1 Labs, Fremont, California, USA}

\author{A. K. Mitchell}

\affiliation{Centre for Quantum Engineering, Science, and Technology (C-QuEST), Belfield, Ireland}

\affiliation{School of Physics, University College Dublin, Belfield, Ireland}

\author{M. Asker}

\affiliation{Equal1 Labs, Fremont, California, USA}

\author{D. Leipold}

\affiliation{Equal1 Labs, Fremont, California, USA}

\author{R. B. Staszewski}

\affiliation{Centre for Quantum Engineering, Science, and Technology (C-QuEST), Belfield, Ireland}

\affiliation{School of Electrical and Electronic Engineering, University College Dublin, Belfield, Ireland}

\affiliation{Equal1 Labs, Nexus UCD, Belfield, Ireland}

\author{E. Blokhina}%

\affiliation{Centre for Quantum Engineering, Science, and Technology (C-QuEST), Belfield, Ireland}

\affiliation{School of Electrical and Electronic Engineering, University College Dublin, Belfield, Ireland}

\affiliation{Equal1 Labs, Nexus UCD, Belfield, Ireland}


\date{\today}

\begin{abstract}
We present an in-depth analysis of a single-electron box (SEB) biased through  a floating node technique that is common in charge-coupled devices (CCDs). The device is analyzed and characterized in the context of single-electron charge-sensing techniques for integrated silicon quantum dots (QD). The unique aspect of our SEB design  is the incorporation of a metallic floating node, strategically employed for sensing and precise injection of electrons into an electrostatically formed QD. To analyse the SEB, we propose an extended multi-orbital Anderson impurity model (MOAIM), adapted to our nanoscale SEB system, that is used to predict theoretically the behaviour of the SEB in the context of a charge-sensing application. The validation of the model and the sensing technique has been carried out on a QD fabricated in a fully depleted silicon-on-insulator (FD-SOI) process on a 22-nm CMOS technology node.
We demonstrate the MOAIM's efficacy in predicting the observed electronic behavior and elucidating the complex electron dynamics and correlations in the SEB. The results of our study  reinforce the versatility and precision of the model in the realm of nanoelectronics and highlight the practical utility of the metallic floating node as a mechanism for charge injection and detection in integrated QDs. Finally, we identify the limitations of our model in capturing higher-order effects observed in our measurements and propose future outlook to reconcile some of these discrepancies.


\end{abstract}

\maketitle

Single-electron boxes (SEB) are a type of nanoscale electronic devices comprising a quantum dot (QD) coupled to a metallic lead through a tunneling junction~\cite{nazarov_blanter_2009}. The chemical potential of the QD is controlled by a gate, coupled capacitively to it (implying no current flowing from the gate to the QD). The distinctive properties of single-electron boxes emerge as a consequence of their nanoscale dimensions, where the quantum nature of charge carriers becomes pronounced, giving rise to phenomena such as Coulomb blockade and quantum tunneling, especially at low temperatures~\cite{Yadav2022-yi,IEEE}.

Quantum properties of SEBs make them extremely charge-sensitive, and these devices can be fabricated with precision engineering. In the light of the immense progress in semiconductor qubit technologies~\cite{anders2023cmos}, there is an increasing interest in sensitive electrometers for silicon spin qubits that would take up a small area and be compatible with large-scale integration. SEB electrometers have been proposed to be used as charge sensors for qubits~\cite{bashir2020single,seb,power2022modelling, PhysRevX.13.011023,mihailescu2023multiparameter} or even as quantum thermometers~\cite{PhysRevA.107.042614}. For that reason, probing the state of a SEB device made on a commercial process will be of particularly great benefit for developing quantum sensing applications. 


In this paper, we present a SEB which is formed by a metallic node (lead) coupled to an electrostatically formed semiconductor QD. The key feature of this study is that the biasing and detecting of the SEB charge states is carried out through a scheme that is inspired by a technique common to the output stage of charge-coupled devices (CCDs)~\cite{janesick1990new}. It is worth noting that CCDs might also be addressing the problem of single-electron detection but in the context of digital imaging~\cite{abramoff2019sensei,tiffenberg2017single}.
Being essentially large-scale integrated systems, CCDs are particularly compatible with commercial and non-commercial semiconductor processes, and the measurement in CCDs are carried out in the charge domain. The feasibility of such biasing schemes has been demonstrated in Ref.~\cite{bashir2020single}. Moreover, quantum nanoelectronics devices lithographically defined in semi-conducting 2D electron gas (2DEG) structures have been integrated recently with charge sensors to measure entropy changes in QD devices~\cite{Andrew1, Andrew2}.

The device is implemented in a commercial fully depleted silicon-on-insulator process on a 22-nm CMOS technology node by GlobalFoundries.  Since the QD of the SEB is controlled electrostatically, we are interested in deriving its quantum mechanical model taking into account effective orbitals and potential shape formation fluctuations and asymmetries which are the aspects of realistic QDs.  For this reason, we develop a type of multi-orbital Anderson impurity model~\cite{Hewson_1993} (MOAIM) to predict the observed voltage from the SEB under the CCD biasing and measurement scheme.  We then compare the SEB experimental characterization at a temperature of  3.5~K with the model and show that the SEB responds to individual charge transitions. This illustrates that it is possible to utilize SEB-CCD electrometers in integrated semiconductor QDs.

\begin{figure*}[t]
        \includegraphics[width=0.75\linewidth]{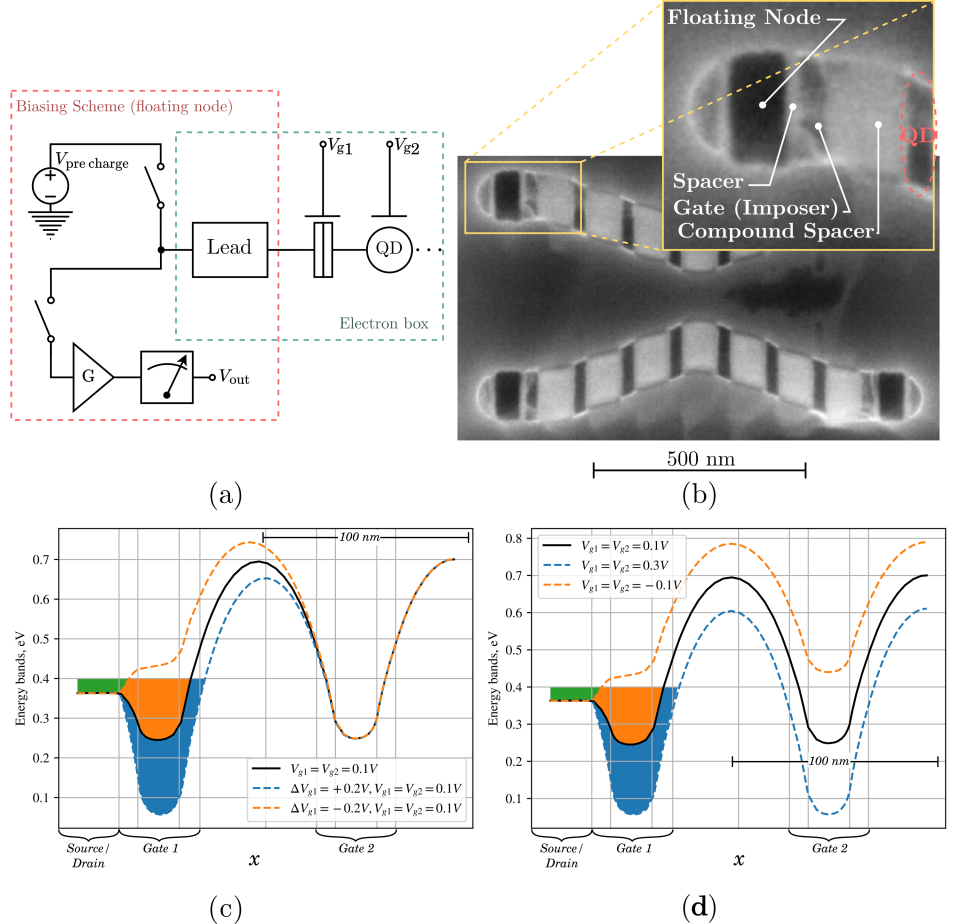}
         \caption{\justifying \textbf{(a)} Schematic of the SEB device with its biasing scheme. 
         \textbf{(b)} Scanning electron microscope (SEM) top-view image of sample~B (more in the next figure). Inset picture shows the part of the device we are utilizing. The floating node is the electron reservoir (lead) coupled via a tunneling barrier to the QD. The tunneling barrier is effectively regulated by \textit{Gate 1}; \textit{Gate 2} on its right is used to form the QD and to control its electrochemical potential. \textbf{(c,d)} QTCAD~\cite{QTCAD} self-consistent simulation of the conduction band, charge density and Fermi level. On the horizontal, we plot the distance $x~$(nm) and on the vertical axis the energy $\mathcal{E}~$(eV). One well (colored) is formed under Gate 1 adjacent to the lead; it is filled with electrons and forms an extended reservoir. The SEB well is formed under Gate~2. Raising $V_{g_1}$ lowers the tunneling barrier between the lead and the SEB QD.  Subfigure~\textbf{(c)} shows the tunneling barrier change with $V_{g_1}$. $V_{g_2}$ manipulates the depth of the QD well. Subfigure~\textbf{(d)} shows the QD depth change with $V_{g_2}$.}
        \label{fig:str}
\end{figure*}

The system is presented in Fig.~\ref{fig:str}. The scheme of the SEB and its biasing principle are illustrated in Fig.~\ref{fig:str}(a). The SEB consists of an electrostatically defined semiconductor QD whose chemical potential is controlled by the gate voltage $V_{\rm g2}$. There is a (metal) lead whose  Fermi energy $E_F$ is lying above the edge of its conduction band. The lead is connected to the QD through a tunneling junction. The Fermi energy of the lead is controlled by the biasing circuit in the following way. A switch is activated to connect the lead to a voltage source that elevates the potential of the lead to some $V_{\text {pre-charge}}$. The switch is deactivated after this, and the node is disconnected from the voltage source --- it is in a ``floating'' state where any change in the number of electrons will result in a significant change of its electric potential due to its small capacitance. Next, the voltage $V_{\rm g1}$ at the gate terminal is adjusted to tune the alignment of the Fermi energy of the lead to the chemical potential of the dot, allowing the tunneling of an electron from the lead to the QD. The second switch is then activated to measure if the electric potential of the lead has changed compared to the original state due to an electron tunneling to the QD. The capacitance of the lead is estimated (via parasitic extraction) to be $0.8$~fF. The target design gain of the voltage amplifier is $80$ (the gain is subject to process variability). This results in an expected $\sim$16~mV step at the output of the voltage amplifier per one electron removed from the lead to the QD. 
The temperature of the set-up is $3.5$~K.

The SEM image of the lead and the QD is shown in Fig.~\ref{fig:str}(b). While it is a part of a large QD array, we apply low voltages  at all the gates. This results  in a very large potential energy barrier separating the quantum dots from each other.  We then can control the tunnelling junction between the reservoir of electrons (lead) and the first quantum dot, while keeping the other dots isolated. 
The parameters of the devices are given in the figure caption. 

The top view of the system is shown in Fig.~\ref{fig:str}(b). The thin Si-film results in a transversal confinement causing a 2DEG behavior of electrons in the film and the in-plane confinement is controlled by the gates. In this class of devices, the formation of wells of the conduction band is defined electrostatically. The dot can form either below the gates or in the spaces between the gates depending on $V_{g_j}$ and other controlled voltages (such as the common-mode voltage $V_{cm}$). 
In the test presented in this paper, at $V_{\text{pre-charge}} = - 410$~mV (after the pre-charge stage, the lead acquires a negative potential) and $V_{g_j}$ ranging up to $0.3$~V, a shallow well is formed below the gates (except for the very first one, adjacent to the lead that forms an extended electron reservoir due to diffusion on electrons). The self-consistent simulation of the conduction band and charge carrier density at 3.5~K,  presented in Fig.~\ref{fig:str}(c) and (d), confirms this assumption. (A single-gate test device was also tested by the means of transport measurement to confirm this.) In this study, we present the results of two different samples that have some variations in the shape of the lead, the spacer and the gate.

In order to perform QD injection, spectroscopy and characterization of its underlying physics, we firstly need to build a theoretical quantum model that can describe and predict the system's behavior. For the purpose of capturing the physical interactions and factors that contribute to the many-body dynamics of the QD, we employ an extended Fermi-Hubbard model for $N$ effective quantum orbitals. The quantum model that describes the QD is expressed by the following Hamiltonian:
\begin{align}
\nonumber H_{QD}&= \sum_{i=1}^{N}\,\sum_{\sigma\in\{\uparrow,\downarrow\}} \epsilon_{i\sigma}\,n_{i\sigma} + \sum_{i\leq j}\,\, \sum_{
\sigma,\sigma'}U_{ij,\sigma\sigma'}\,n_{i\sigma}\,n_{j\sigma'}
\end{align}
\noindent where $\epsilon_{i\sigma}$ is the on-site potential for orbital $i$,  $U_{ij,\sigma\sigma'}$ is the electrostatic Coulomb coupling between an electron at orbital $i$ and spin $\sigma$ and another electron at orbital $j$ and spin $\sigma'$ with $(i,\sigma) \neq (j, \sigma')$,
$c_{i\sigma}^{\left(\dagger\right)}$ is the Fock space fermionic annihilation (creation) operator for a fermion at orbital $i$ and spin $\sigma$ and $n_{i\sigma}= c_{i\sigma}^\dagger\,c_{i\sigma}$ is the number operator.
The creation/annihilation operators satisfy the fermionic algebra anti-commutation relations $\{c^{\dagger}_{j\sigma},c_{k\sigma'}\}=\delta_{jk}\, \delta_{\sigma \sigma'}$ and $\{c^\dagger_{j\sigma}, c^\dagger_{k\sigma'}\} = \{c_{j\sigma}, c_{k\sigma'}\} = 0$ and act upon the system's Fock space; for $\mathcal{M}$ fermions it is defined as $\mathfrak{F}_{\mathcal{M}}\equiv \bigoplus_{m=0}^{\mathcal{M}} \mathcal{H}_m$, with $\mathcal{H}_m\equiv \mathcal{H}^{\otimes m}$ being the $m-$fermion Hilbert subspace.
\\

\begin{figure*}[t]
     \centering
        \includegraphics[width=\textwidth]{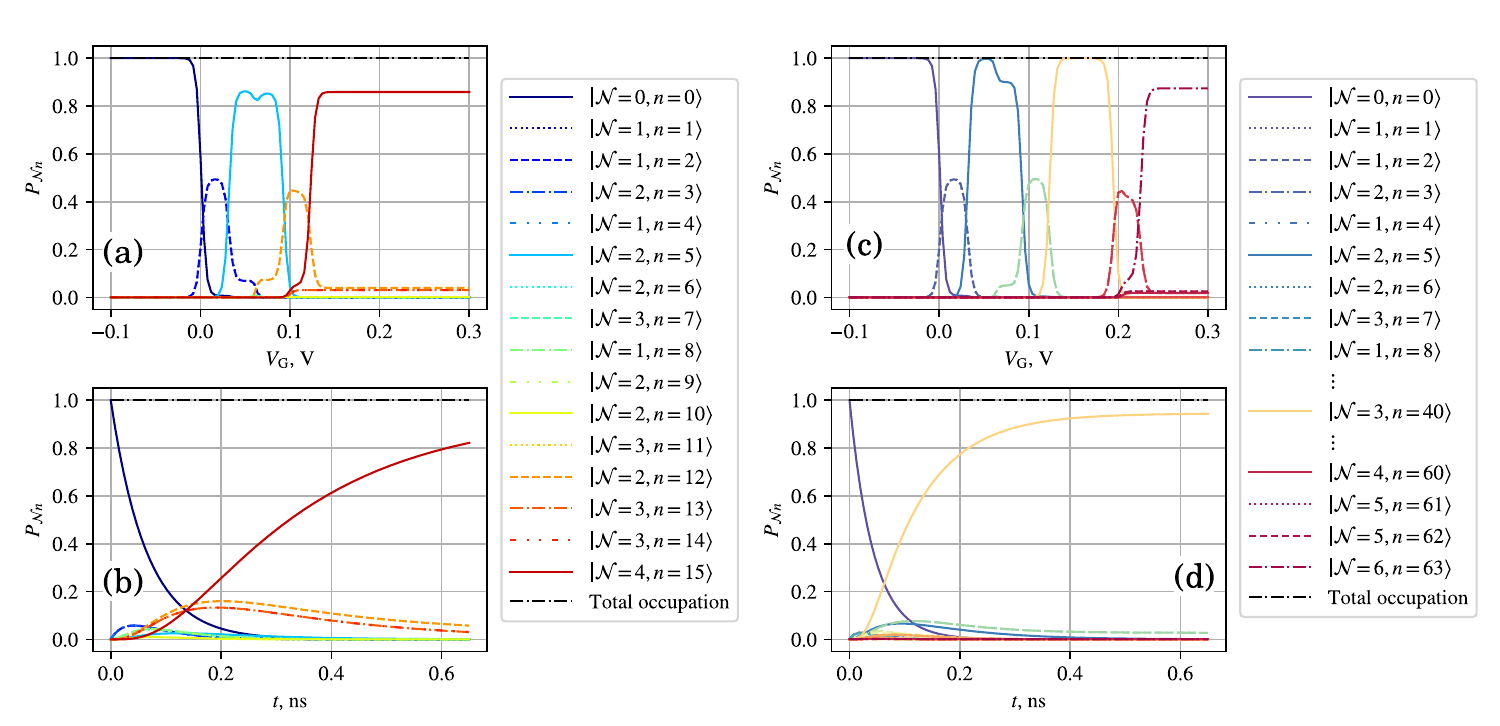}
         \caption{\justifying{\bf Fock state occupancy probabilities $P_{\mathcal{N}n}$ at $T=3.5$~K and hybridization energy $\tau_{k\sigma} = 20~\mu$eV.} Probability plots as functions of different system parameters. \textbf{(a)} Fock states' occupancies $P_{\mathcal{N}n}$ as a function of applied voltage $V_g$ sweep for sample A for $t\approx 0.6$~ns.
         \textbf{(b)} Dynamical evolution of Fock states occupancies as a function of time, for $V_g \approx 0.18$~V for sample A. 
         \textbf{(c)} Same as \textbf{(a)} but for the model corresponding to sample B. We see more plots, because we consider $N=3$ orbitals in the model for this sample, so we can have up to a maximum of $\mathcal{N}_{max} = 6$ electrons present in the QD and a total of $64$ states. \textbf{(d)} Same as \textbf{(b)} but for the model corresponding to sample B.}
        \label{popbias}
\end{figure*}

For the physical specifications of our system we have taken the QD to be of volume $v_{\rm QD} = (80 \times 30 \times 6.25)$~nm$^3$  and to be composed of $N=2$ effective orbitals for sample A and $N=3$ for sample B; these are chosen phenomenologically and ad-hoc. The dimensions of the dot are taken from the physical parameters of the system, and will provide a very good correspondance to the observed energy levels. 
Since  $d^{(z)} = 6.25$~nm $\ll\, d^{(x)},d^{(y)}$ and so $\Delta\,\epsilon^{(z)} \gg \Delta\,\epsilon^{(x)},\Delta\,\epsilon^{(y)}$, we restrict ourselves to the orbitals formed due to confinement in the smaller $\hat{x},\hat{y}-$dimensions. The contributing energies are the Coulomb intra and inter-orbital coupling energies $U_{ij \,,\, \sigma\sigma'} = 3.1$~meV (Si),
and the \textit{effective} on-site confinement energies $\epsilon_{i\sigma}^{\,A}\in\{1.56, 5.63\}$~(meV) and $\epsilon_{i\sigma}^{\,B}\in\{1.56, 5.63, 10.91\}$~(meV), with $\epsilon_{i\sigma}<\epsilon_{j\sigma}$ for $i<j$. These are derived from a symmetric finite quantum well calculation, fine-tuned by a common-mode voltage $V_{cm} \approx -427$~mV (which is close to the $-440$~mV used in QTCAD) and since our formed QD need not be completely symmetrical, we add a random fluctuation of the energy levels to account for potential imperfections in our electrostatically formed QD. That is, we have $\epsilon_{i\sigma}^{X} = \epsilon_{k\sigma}^{X\,(ideal)} + \delta\epsilon^{X\,(asym)}_{k\sigma}$, with $\delta \epsilon_{k\sigma}^{ X\,(asym)}$ sampled from a normal distribution $\mathcal{N}(\mu,\sigma^2)$ with mean $\mu = 0$ and standard deviation $\sigma = 20\%$. 
Finally we utilize the transversal and longitudinal electron masses~\cite{Kittel2004-rm} as $m_t = 0.19 m_e$ (Si) and $m_l = 0.98 m_e$ (Si), respectively. 

\begin{table*}[htb]
\caption{\justifying {\bf Table of the allowed Fock eigenstates $\ket{\mathcal{M},m}$ of the QD Hamiltonian $H_{QD}$ of each Fock subspace.} The states' coefficients $\alpha_{mk},\beta_{mk},\ldots,$ depend on the parameters of the system $\epsilon_{i\sigma},\,U_{jk,\sigma\sigma'}$ which are in turn dependent on the physical architecture and characteristics of the QDs and control voltages $v_{QD}\,,\,V_g\,,\,\Delta V_{g_1}$; we include here the compact expressions, since the full ones are very complex and long to write out explicitly. The states are denoted as $\ket{s_1;s_2;\, \ldots \,;s_{N}}\equiv \bigotimes_{n=1}^N\ket{s_n}$, with $s_k$ the $\hat{z}-$spin projection of the $k^{th}-$QD level occupation. We label with $m_i^*$ the maximum value of $m_i$ for $\mathfrak{F}_i$ and with $m^*$ its overall maximum value.\label{tab}}
\begin{tabular}{@{}c|c@{}}
\toprule
{\centering Number of electrons $\nu_{e^-}$ }&{\centering Fock space eigenstate $\ket{\mathcal{M},m}$ }\\[3pt] 
\hline
&\\[-2pt]
$0$ & $\ket{0,1}= \ket{0;0;\,\ldots\,;0}$\\[3pt]
$1$ & $\ket{1,m_1}= \sum_{k=1}^{m_1}\alpha_{m_1k}\ket{s_{1_{m_1k}};s_{2_{m_1k}};\,\ldots\,;s_{N_{m_1k}}},\quad m_1\in\left\{1,...,\footnotesize\begin{pmatrix} \mathcal{M}_{max} \\ 1 \end{pmatrix}\right\}$\\[8pt]
$2$ & $\ket{2,m_2}= \sum_{k=1}^{m_2}\beta_{m_2k}\ket{s_{1_{m_2k}};s_{2_{m_2k}};\,\ldots\,;s_{N_{m_2k}}},\quad  m_2\in\left\{m_1^*+1,...,\footnotesize m_1^*+\begin{pmatrix} \mathcal{M}_{max} \\ 2 \end{pmatrix}\right\}$\\[8pt] 
\LARGE\vdots & \quad \quad \LARGE\vdots \\[8pt]
$\mathcal{M}_{max}$ & $\ket{\mathcal{M}_{max},m^*}=\ket{\uparrow\downarrow;\uparrow\downarrow;\,\ldots\,;\uparrow\downarrow}$\\[8pt]
\botrule
\end{tabular}
\end{table*}

In addition to the above Hamiltonian which describes the dynamics of the  isolated QD, we employ a lead Hamiltonian $H_{\rm lead}$, which corresponds to the metallic floating node in the actual structure.  We treat it as a semi-classical reservoir of electrons, from which they can jump in and out, with different energies $\{\lambda_{r\sigma}\}$ and a Fermi level $E_F$ controlled externally through manipulation of a tunable electrochemical potential $\mu_{ch}$. We include only a left lead ($\mathcal{L}$) coupled to the QD. In addition, we add the coupling between the lead and the QD (hybridization), so the total Hamiltonian for our SEB will have the form of a MOAIM:
\begin{equation}
H_{sys} = H_{lead} \,+\,H_{QD}\,+\sum_{r\,\sigma\,\in\,\mathcal{L}}\,\left(\tau_{r\,\sigma}\,w_{r\,\sigma}^\dagger\,c_{\sigma}\,+\,h.c.\right)
\end{equation} 
with $H_{\rm lead} = \sum_{r\,\sigma\,\in\,\mathcal{L}}\,\lambda_{r\,\sigma}\,w_{r\,\sigma}^\dagger\,w_{r\,\sigma}$ with $\lambda_{r\sigma}$ the energy of the $r$ level and spin $\sigma$ of the metallic node Hamiltonian, $w_{r\,\sigma}^{(\dagger)}$ the fermionic annihilation (creation) operator of an electron of energy $r$ level and spin $\sigma$ in the lead, $\tau_{r\,\sigma}$ the hybridization (coupling) energy between each lead level and the QD and $h.c.$ is the Hermitian conjugate counterterm.

We consider a uniform hybridization $\tau_{r\sigma} = \tau$ which is also manipulated via gate voltages $\Delta V_{g_1}$ in the device. In our simulation, its values range in the $\tau_{r\sigma}$ = 3--40\,$\mu$eV regime, which satisfies $\tau_{r\sigma} \ll \mathcal{E}$ for any energy scale $\mathcal{E}$ of our system. 
The full Hamiltonian $H_{sys}$ can then be used to model the quantum transport properties of the system, and how the QD electronic occupation depends on applied potentials \cite{kouwenhoven1997electron,shangguan2001quantum,gurvitz1998rate}.

We treat the QD in $\mathfrak{F}_{\mathcal{M}_{max}}$ Fock space, allowing up to $\mathcal{M}_{max}=4$ and $\mathcal{M}_{max}=6$ electrons to occupy the structure for sample A and sample B, respectively.
Our density matrix $\rho=\left\{Q_{\mathcal{MM}'}^{\,mm'}\right\}_\mathcal{MM'}^{\,mm'}$, with $Q_{\mathcal{MM'}}^{\,mm'}=\ket{\mathcal{M}\,,\,m}\bra{\mathcal{M'}\,,\,m'}$, the Hubbard operator has diagonal elements $P_{\mathcal{M}\,m}$, with $\sum_{\mathcal{M},m} P_{\mathcal{M}\,m}=1$, and we show them symbolically in Table~\ref{tab}. Each of the Fock eigenstates with $\mathcal{M}$ electrons will be a superposition of $m$ states, with $m=\begin{pmatrix} \mathcal{M}_{max} \\ \mathcal{M} \end{pmatrix}=$ \scalebox{1.3}{$\frac{\mathcal{M}_{max}!}{\left(\mathcal{M}_{max}-\mathcal{M}\right)!\,\mathcal{M}!}$}.

We assume that the lead is weakly coupled to the QD and therefore keep up to second-order hybridization terms $\mathcal{O}(\tau_{k\sigma}^2)$, so we can ignore off-diagonal elements in $\rho$ and significant mixing of Fock space states.
Consequently, the dynamical evolution of the population numbers $P_{\mathcal{M}\,m}=P_{\mathcal{M}\,m}(t)$ is simplified to a set of partial differential master equations~\cite{PhysRevB.73.205333}:
\beq
\frac{\partial P_{00}}{\partial\,t}=-\frac{1}{\hbar}\,\sum_{m=1}^{\mathcal{M}_{max}}\,\Gamma^{\,res}_{00\,,\,1m}\left[\bar{n}_{\,res}^{\,+}\,\left(\Delta_{1m\,,\,00}\right) P_{00}- \bar{n}_{\,res}^{\,-}\left(\Delta_{1m\,,\,00}\right)P_{1m}\right] \eeq
\begin{align} \nonumber &\frac{\partial P_{\mathcal{M}m}}{\partial\,t}=\frac{1}{\hbar}\,\sum_{m'}\Big\{\,\Gamma^{\,res}_{\mathcal{M}-1\,m'\,,\,\mathcal{M}m}\big[\bar{n}_{\,res}^{\,+}\,\left(\Delta_{\mathcal{M}m\,,\,\mathcal{M}-1\,n'}\right)P_{\mathcal{M}-1\,m'}\\
 &-\bar{n}_{\,res}^{\,-}\left(\Delta_{\mathcal{M}m\,,\,\mathcal{M}-1\,m'}\right)P_{\mathcal{M}m}\big]-
\Gamma^{\,res}_{\mathcal{M}m\,,\,\mathcal{M}+1\,n'}\big[\bar{n}_{\,res}^{\,+}\,\left(\Delta_{\mathcal{M}+1\,m'\,,\,\mathcal{M}m}\right)\nonumber \\
&\times P_{\mathcal{M}m}- \bar{n}_{\,res}^{\,-}\left(\Delta_{\mathcal{M}+1\,n'\,,\,\mathcal{M}m}\right)P_{\mathcal{M}+1\,m'}\big]\Big\}\end{align}
\begin{align}
&\nonumber \frac{\partial P_{\mathcal{M}_{max}\,m^*}}{\partial\,t}=\frac{1}{\hbar}\,\sum_{m=1}^{\mathcal{M}_{max}}\,\Gamma^{\,res}_{\mathcal{M}_{max}-1\,m\,,\,\mathcal{M}_{max}\,m^*}\big[\bar{n}_{\,res}^{\,+}\,\left(\Delta_{\mathcal{M}_{max}\,m^*\,,\,\mathcal{M}_{max}-1\,m}\right)\\ &
\times P_{\mathcal{M}_{max}-1\,m} -\bar{n}_{\,res}^{\,-}\left(\Delta_{\mathcal{M}_{max}\,m^*\,,\,\mathcal{M}_{max}-1\,m}\right)P_{\mathcal{M}_{max}\,m^*}\big]
\end{align}
\\
where $\Gamma^{\,res}_{\mathcal{M}m\,,\,\mathcal{M}'\,m'}=2\pi\,\sum_{r\sigma\,\in\,\mathcal{L}} |\tau_{r\sigma} \alpha_{\mathcal{MM'}}^{\,mm'}|^2\, \delta\big(\Delta_{\mathcal{M'}\,m'\,,\,\mathcal{M}m}-$ $\lambda_{r\sigma}\big)$ is the tunneling rate between the metallic lead and the QD with $\Delta_{\mathcal{M}m\,,\,\mathcal{M'}m'} \equiv E_{\mathcal{M}m} - E_{\mathcal{M'}m'}$ is the energy difference between the two Fock eigenstates, $\alpha_{\mathcal{MM'}}^{\,mm'} \equiv \bra{\mathcal{M},m}\,c_{\sigma}\ket{\mathcal{M'},m'}$ are the transition elements, $\delta(x-\epsilon_{r\sigma})$ is the Dirac delta function, $\bar{n}_{\,res}^{\,+}(x)=\frac{1}{e^{(x-\mu_{ch})/{k_B\,T}}+1}$ is the Fermi-Dirac statistical function and $\bar{n}_{\,res}^{\,-}(x)=1-\bar{n}_{\,res}^{\,+}(x)$.
\\
\begin{figure*}[t]
     \centering
        \includegraphics[width = 0.85\linewidth]{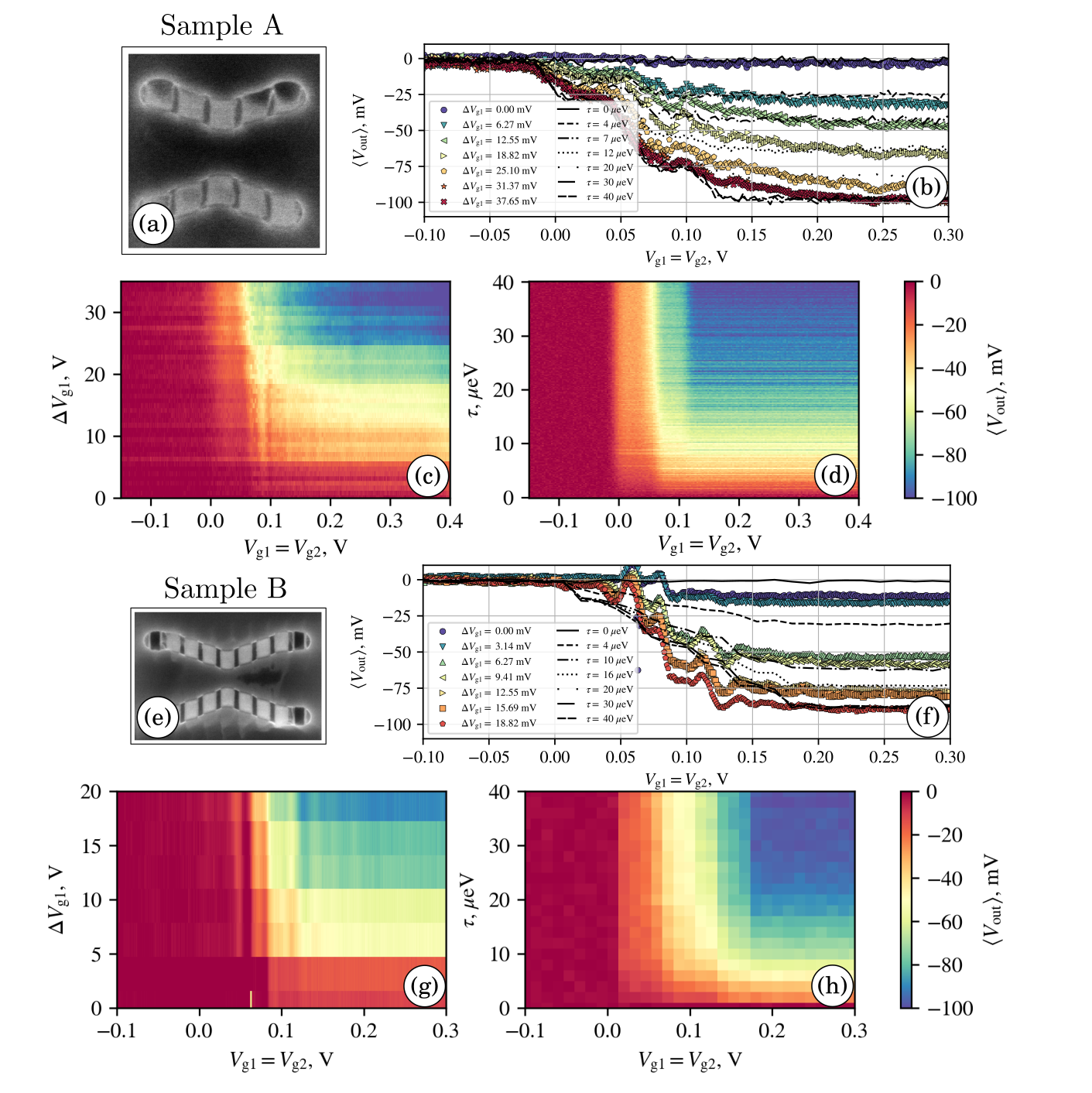}
        \caption{\justifying{\bf Injections line-plots and heatmaps for two different structures, labeled sample A and sample B.} \textbf{(a)} Sample A architecture. \textbf{(b)} Lineplots of experimental measurements (colored dots) and theoretical predictions (black dashed lines) for sample A. \textbf{(c)} Heatmap produced from the experiment of sample A structure. \textbf{(d)} Heatmap produced from our MOAIM for the same structure; in the transversal axis we have left the hybridization energy units from the model. \textbf{(e)} Sample B architecture. \textbf{(f)} Lineplots of experimental measurements (colored dots) and theoretical predictions (black dashed lines). Again we have left the model hybridization energies in their original expressions. \textbf{(g)} Heatmap produced from experimental measurements for sample B. \textbf{(h)} Heatmap produced from our MOAIM for the same structure.}
        \label{avgnum}
\end{figure*}

Initializing our system with no electrons in it, $P_{00}(t=0)=1$, we show the dependence of the population numbers close to equilibrium (steady state), with respect to $\mu_{ch}$ of the lead in Fig.~\ref{popbias}(a) for sample A and Fig.~\ref{popbias}(c) for sample B, respectively. 
We can see the dynamic evolution of the Fock states to approach thermal equilibrium in Fig.~\ref{popbias}(b) for sample A and in Fig.~\ref{popbias}(d) for sample B.
Given some initial conditions, there is a clear threshold electrochemical lead potential (i.e. gate voltage $V_g$) $\mu_{ch,\,\rm th}$ for which we have a many-body stochastic injection in the structure which is related to the energy level spacing between the lead and the first non-trivial Fock state of the QD. Using the population numbers and the corresponding electron number $\nu(\mu_{ch})$ of each Fock state, we can compute an average:
\begin{equation}\left\langle \nu_T(\mu_{ch}) \right\rangle \equiv \mathbb{E}\left[\nu_T(\mu_{ch})\right] = \mkern-12mu \sum_{\mathcal{M}=0}^{\mathcal{M}_{max}}\sum_m^{m_{max}} \nu_{\mathcal{M}m}(\mu_{ch}, T)\times P_{\mathcal{M}m}(\mu_{ch}, T)
\end{equation}
where $\nu_{\mathcal{M}m}(\mu_{ch},T)$ and $ P_{\mathcal{M}m}(\mu_{ch},T)$ is the number of electrons and occupational number of state $\ket{\mathcal{M},m}$ in the QD for an electrochemical potential $\mu_{ch}$ and temperature $T$, respectively.
 
We can connect $\nu(\mu_{ch})$ to the actual voltage measurements in our detectors using the transformation functions $\langle V^{A}_{out} \rangle \approx -23\,\langle \nu_T \rangle$~mV and $\langle V^{B}_{out} \rangle \approx -16\,\langle \nu_T \rangle$~mV, where $V^{X}_{out}$ is the experimentally measured voltage on structure $X$, as we show in Fig.~\ref{avgnum}.
Here, we plot the obtained $\langle V^{X}_{out} \rangle$ as a function of the applied gate voltage $V_g$ in order to describe approximately the measured voltage drop. We use the transformation function $V_g = 8\,\mu_{ch}\, (V/eV)$ obtained from QTCAD simulations. Moreover, to account for the effects of noise in the measurement, we use a Gaussian noise filter in our calculated output observable. That is we plot: $\langle \tilde{V}^{X}_{out} \rangle = \langle V^{X}_{out} \rangle \times \mathcal{G}_{noise}$, where $\mathcal{G}_{noise} \in \mathcal{N}(\mu, \sigma^2)$ with $\mu = 1$ and $\sigma = 4\%$, for both structures. Sweeping over $V_{\rm g1}$ and $\Delta V_{g_1}$ voltages in the experiment corresponds to sweeping over $\mu_{ch}$ and $\tau$ in our model.

As a conclusion, we would like to highlight some key points of this work. Firstly, we showed that the resolution of our device is sensitive to single-electron injection within some variance induced by thermal noise due to the finite operating temperature. This shows that both the device and the incorporation of a metallic floating node with standard CCD circuitry are efficient in relevant quantum charge sensing applications. Moreover, the agreement between our simulations from QTCAD and MOAIM with the experimentally measured data further hints at the device behaving as an SEB and a QD forming under Gate 2. The predictions of the MOAIM are compatible both with the applied voltages and the QD geometry in the actual device and it captures effectively the most significant aspects of the experiment. These are the measured voltage at electron injection plateaus (quantized charge, Coulomb blockade), the stretching of the curves when the coupling between metallic node and the QD is varied $\left(\partial \langle V^{X}_{out} \rangle / \partial \tau > 0\right)$ and the average decrease in the variation between curve gaps $\left(\partial^2 \langle V^{X}_{out} \rangle / \partial \tau^2 < 0\right)$.

On the other hand, there are some potentially interesting physics that our model does not capture. Some of these are the small "S" shaped bumps in the experimental data which are located where injection happens (more apparent in sample B) and
the voltage-dependent drift in the experimental curves (more apparent in sample A). Finally, one can in principle go beyond the semi-classical rate-equation treatment presented above to include QD-lead entanglement effects~\cite{minarelli2022linear}, renormalization phenomena at low-temperatures, and spin-flip scattering producing more subtle quantum behavior such as the Kondo effect \cite{mitchell2011two}.  We leave all of the aforementioned as interesting outlooks for future research.

\section*{Author Declarations}
\subsection*{Conflict of interest}
The authors have no conflicts to disclose.

\section*{Data availability}
The data that support the findings of this study are available from
the corresponding authors upon reasonable request.

\bibliography{bibliography}

\end{document}